\documentclass[12pt]{article}

\catcode`\@=11
\@addtoreset{equation}{section}

\global\arraycolsep=2pt
\def\fnote#1#2{\begingroup\def\thefootnote{#1}\footnote{#2}\addtocounter
{footnote}{-1}\endgroup}

\usepackage{amsbsy,amssymb,latexsym,amsfonts,amsmath,cite}
\usepackage{graphicx,color}

\newcommand\CW{{\mathcal W}}
\newcommand\CL{{\mathcal L}}

\newcommand\e{\mathrm{e}}
\newcommand\diag{\mathrm{diag}}

\newcommand{\adss}[2]{{AdS$_{#1}\times$S$^{#2}$}}
\newcommand{\ads}[1]{{AdS$_{#1}$}}
\newcommand{\s}[1]{{S$^{#1}$}}

\newcommand\Tr{\mathrm{Tr}}

\begin{document}

\begin{flushright}
\parbox{4.2cm}{
OIQP-07-09\hfill \\
NSF-KITP-07-177 \hfill 
}
\end{flushright}

\vspace*{0.5cm}

\begin{center}
{\bf\Large
A Semiclassical String Description of \\ Wilson Loop 
with Local Operators
}

\vspace*{1.5cm}
{\large 
Makoto Sakaguchi$^a$\fnote{$\ddagger$}{e-mail: 
\texttt{makoto\_sakaguchi@pref.okayama.jp}}\quad and \quad
Kentaroh Yoshida$^b$\fnote{$*$}{e-mail: \texttt{kyoshida@kitp.ucsb.edu}}
}
\vspace*{1.5cm} \\ 
$^a$ {\it Okayama Institute for Quantum Physics\\
1-9-1 Kyoyama, Okayama 700-0015, Japan}

\vspace*{0.5cm}

$^b$ {\it Kavli Institute for Theoretical Physics, \\ 
University of California,  \\
Santa Barbara,  CA.\ 93106, USA }
\end{center}

\vspace*{1cm}

\begin{abstract}

We discuss a semiclassical string description to circular Wilson loops
without/with local operator insertions. Type IIB string theory on
AdS$_5\times$S$^5$ is expanded around the corresponding classical
solutions with respect to fluctuations and semiclassical quadratic
actions are computed. Then the dual corresponding operators describing
the fluctuations are discussed from the point of view of a small
deformation of the Wilson loops. The result gives new evidence for
AdS/CFT correspondence. 

\end{abstract}

\newpage

\section{Introduction}

Almost a decade has passed from a discovery of AdS/CFT correspondence
\cite{AdS/CFT:M, AdS/CFT:GKPW}. Now it is firmly supported by enough
 evidences, but there is no proof
of it now. 
Hence it is still important to continue to
seek further, new confirmation to support it.

\medskip 

One of the difficulties is to analyze type IIB string on
AdS$_5\times$S$^5$\,. The action is constructed in \cite{MT} and 
its classical integrability is shown in \cite{BPR}. However,   
it still seems difficult to quantize the theory manifestly,  
simply because the action is quite non-linear.  
A sensible way is
to find a solvable subsector such as the BMN sector \cite{BMN}. The BMN
sector is pulled out by taking a Penrose limit \cite{Penrose}. 
Then the simplified string theory is exactly solvable
\cite{Metsaev,MT2}, and hence one can test the duality at stringy level
though the argument is restricted to a certain region.

\medskip 

It is pointed out in \cite{GGK} that a non-relativistic limit of type
IIB string on AdS$_5\times$S$^5$ gives a new arena to test the
AdS/CFT. It is shown in \cite{SY:non-rela2} that the limit is
regarded as a semiclassical approximation around a static AdS$_2$
solution \cite{Wilson} like as the Penrose limit is around a BPS particle
\cite{GKP}. This equivalence holds even for AdS-branes
\cite{SY:non-rela, SY:non-rela2}. With this semiclassical
interpretation it has been shown that the corresponding operator in the
gauge theory is nothing but a small deformation of straight Wilson line \cite{SY:non-rela2}.

\medskip 

The purpose of this paper is to generalize the result for the straight
line to circular Wilson loops without/with local operator insertions. An
AdS$_2$ solution corresponds to a 1/2 BPS circular Wilson loop {\it
without} the insertions \cite{CWL}. A semiclassical approximation around
the solution has already been studied in \cite{DGT}. We newly compute a
quadratic action around the solution corresponding to a circular Wilson
loop {\it with} the local operators, $Z^J$ and its complex conjugate.
Here $Z$ is a complex scalar composed of the two real scalar fields in
$\mathcal{N}$=4 SYM like $Z \equiv \phi_1 + i\phi_2$\,. The resulting
quadratic fluctuations describe the action in \cite{DGT} around
$\sigma=0$ while those behave as a pp-wave string at $\sigma=\infty$\,.

\medskip 

Then we clarify that the fluctuations correspond to a small deformation
of the circular Wilson loops without/with the insertions. In particular,
the dictionary of impurity insertion is derived. With no local operator
the dictionary is the same as in the case of the straight line
\cite{SY:non-rela2}. This result is not so surprising since the
difference between a straight Wilson line and a circular Wilson loop is
the behavior at infinity and it only gives an anomalous contribution to 
the expectation value (and the value of classical action of 
the corresponding string solution). However the local behavior around a finite
point should not be different. With the local operators, it is the same
as the case without them apart from the insertion points while it is
nothing but the BMN dictionary \cite{BMN} on the inserted local
operators. This result nicely agrees with the behavior of the
semiclassical action.

\bigskip 

This paper is organized as follows. In section 2 we reproduce a
semiclassical action around an AdS$_2$ solution whose boundary is a
circular Wilson loop with no local operator. Then, in section 3, we
discuss the corresponding operators in the gauge theory from a small
deformation of the circular Wilson loop. In section 4, as a further
generalization, we consider a semiclassical action around the
Miwa-Yoneya solution \cite{MY}, which is a generalization of the
solution constructed by Drukker-Kawamoto \cite{DK} in the Lorentzian
case. This solution corresponds to a circular Wilson loop with local
operator insertions.  The resulting action interpolates the pp-wave
string action and the semiclassical action around the AdS$_2$ as
expected. In section 5 we consider a small deformation of the Wilson
loop corresponding to the semiclassical action obtained in section 4.
The configuration of the Wilson loop is more involved. Section 6 is
devoted to a summary and discussions.

\section{Semiclassical limit around a circular solution}

In this section, as a warming up, let us consider a classical string
solution whose boundary describes a circular Wilson loop \cite{CWL} {\it
without} local operator insertions. Note that the quadratic string
action with respect to the fluctuations has already been computed by
Drukker-Gross-Tseytlin \cite{DGT}. To make the present paper
self-contained, however, we shall rederive the result of \cite{DGT}
here. Then we show the agreement between the fluctuations around the
classical solution in the string side and those around the circular
Wilson loop in the gauge theory.

\subsection{Classical solution for a circular Wilson loop}

First let us discuss a classical solution describing a circular Wilson loop.  
We begin with the string action in the Polyakov formulation and 
the bosonic part is given by 
\begin{eqnarray}
S_{\rm B}&=&\frac{\sqrt{\lambda}}{4\pi}\int\! d^2\xi\,
\sqrt{\gamma}\gamma^{ij}\partial_iX^M\partial_jX^NG_{MN}~,
\label{string bosonic}
\end{eqnarray}
where $\gamma_{ij}$ is an auxiliary world-sheet metric and we work in
conformal gauge, $\sqrt{\gamma}\gamma^{ij}=\delta^{ij}$.  The spacetime metric
$G_{MN}$ describes AdS$_5\times$S$^5$ and it is given by
\begin{eqnarray}
ds^2=\frac{1}{z^2}(r^2d\theta^2+dr^2
+dx_2^2+dx_3^2
+dz^2)
+d\Omega_5^2~. \nonumber
\end{eqnarray}
Hereafter we will work in Euclidean signature and Poincare
coordinates. See Appendix A for the detail expressions of
vielbeins and spin connections. 

\medskip 

The equation of motion reads
\begin{eqnarray}
0&=& -\partial_i(\delta^{ij}\partial_j X^NG_{MN})
+\frac{1}{2}\partial_MG_{PQ}\delta^{ij}\partial_i X^P\partial_j X^Q~.
\label{EL}
\end{eqnarray}
The Virasoro constraints to be imposed are
\begin{eqnarray}
0&=& G_{MN}(-\dot X^M\dot X^N+ X'{}^M X'{}^N)~,~~~
0=G_{MN}(\dot X^MX'{}^N)~.
\label{Vir}
\end{eqnarray} 
It is easy to see that 
\begin{eqnarray}
z=\tanh\sigma~, \qquad 
r=\frac{1}{\cosh\sigma}~, \qquad 
\theta=\tau
\label{classical sol}
\end{eqnarray}
solves the equation of motion (\ref{EL}) and the Virasoro conditions
(\ref{Vir}). The classical solution (\ref{classical sol}) at $\sigma=0$
describes a circular Wilson loop with a unit radius on the boundary
\begin{eqnarray}
z=0~, \qquad r=1~,  \qquad \theta=\tau~. \nonumber 
\end{eqnarray}

\medskip 

Here it is valuable to comment on the relation between a circular loop
and a straight line. First we move to Cartesian coordinates,
\begin{eqnarray}
z=R\tanh\sigma~, \qquad 
x_0=R\frac{\sin\tau}{\cosh\sigma}~, \qquad 
x_1=R\left(\frac{\cos\tau}{\cosh\sigma}-1\right)~, \nonumber 
\end{eqnarray}
where the radius of the circle $R$ has been recovered. Then let us
rescale $\tau$ and $\sigma$ as
\begin{eqnarray}
\tau\to\frac{\tau}{R}~, \qquad 
\sigma\to\frac{\sigma}{R}  \nonumber 
\end{eqnarray}
and take the large $R$ limit. As a result,  (\ref{classical sol})
is reduced to a static \ads{2} solution
\begin{eqnarray}
z=\sigma~, \qquad 
x_0=\tau~, \qquad 
x_1=0~. \nonumber 
\end{eqnarray}
At the boundary $\sigma=0$ this solution describes a straight Wilson
line.

\subsection{Semiclassical limit}

Next we consider a semiclassical approximation of 
the full type IIB string on AdS$_5\times$S$^5$ around 
the classical solution (\ref{classical sol}). 

Let us expand the string action 
(\ref{string bosonic})
about the classical solution
(\ref{classical sol})
\begin{eqnarray}
&& z=\tanh\sigma +\tilde z~, \qquad 
r=\frac{1}{\cosh\sigma}+\tilde r~, \qquad 
\theta=\tau+\tilde \theta~, \nonumber \\ 
&& x_{2,3}=0+\tilde x_{2,3}~, \qquad 
\varphi_i=0+\tilde\varphi_i \quad (i=1,\ldots,5), \nonumber 
\end{eqnarray}
where quantum fluctuations are denoted as the symbols with tilde like
$\tilde X$\,.  Hereafter 
the overall factor of the action (\ref{string bosonic}),
$\sqrt{\lambda}$ is absorbed into the definition of fluctuations by
rescaling the variables as $\tilde X\to {\lambda^{-1/4}}\tilde X$\,.
Then the value of $\lambda$ should be taken to be large in order for the
semiclassical (quadratic) approximation to be valid.

An additional redefinition of the variables is performed as
\begin{eqnarray}
\bar r=\tilde r\coth\sigma~, \quad 
\bar x=\tilde x\coth\sigma~, \quad 
\bar z=\tilde z\coth\sigma~, \quad 
\bar \theta=\tilde \theta \frac{1}{\sinh\sigma}~,~~ \nonumber 
\end{eqnarray}
and the following quadratic action is obtained,  
\begin{eqnarray}
S_{\rm 2B}&=&\frac{1}{4\pi}\int d^2\xi\Big[
(\partial \bar \theta)^2
+(\partial \bar r)^2
+(\partial \bar x_2)^2
+(\partial \bar x_3)^2
+(\partial \bar z)^2
+(\partial \tilde \varphi_i)^2
\nonumber\\&&
+\frac{2}{\cosh^2\sigma}\bar z^2
-4\frac{\sinh\sigma}{\cosh^2\sigma}\bar z\bar r
+\bar r^2
+\frac{4}{\cosh\sigma}\bar z\bar r'
+4\bar r\dot{\bar\theta}
-\frac{4}{\sinh\sigma}\bar z\dot{\bar\theta}
~~\Big]\,. 
\label{fluc}
\end{eqnarray}

\medskip 

Although one should impose the Virasoro constraints to eliminate the
longitudinal modes, it is not an easy task for small fluctuations. Thus
we shall take another course following \cite{DGT} instead. Note that the
action (\ref{fluc}) can be rewritten with conformal gauge as follows:
\begin{eqnarray}
S_{\rm 2B}&=&
\frac{1}{4\pi}\int \!d^2\xi 
\Big[
\delta^{ij}D_i\zeta^AD_j\zeta^B\delta_{AB}
+X_{AB}\zeta^A\zeta^B
\Big]\,, \label{fluc 2} 
\\ && 
\qquad \zeta^A=(\bar\theta,\bar r, \bar x_2,\bar x_3,
\bar z; \bar\varphi_i)\,. \nonumber 
\end{eqnarray}
Here the following quantities have been introduced: 
\begin{eqnarray}
D_i\zeta^A&=&\partial_i\zeta^A+\Omega^A_i{}_B\zeta^B\,, \quad 
\Omega^A_i{}_B=\partial_iX^M \Omega^A_M{}_B\,, \quad E^A_i=\partial_iX^M
E_M^A\,, 
\nonumber \\
X^{ab} &=&
\delta^{ab}\delta^{ij}E^c_iE^d_j\delta_{cd}
-\delta^{ij}E^a_iE^b_j~,~~~ \nonumber \\  
X^{a'b'} &=&
-\delta^{a'b'}\delta^{ij}E^{c'}_iE^{d'}_j\delta_{c'd'}
+\delta^{ij}E^{a'}_iE^{b'}_j~, \nonumber 
\end{eqnarray}
where $E^A$ and $\Omega^A{}_B$ are vielbein and spin connection of
\adss{5}{5} evaluated with the classical solution (\ref{classical sol}).
In the present case we obtain
\begin{eqnarray}
&&
D_i\zeta^0=\left(\partial_\tau\zeta^0-\frac{1}{s}\zeta^4+\zeta^1,
\partial_\sigma\zeta^0 \right)\,, \qquad 
D_i\zeta^1=\left(\partial_\tau\zeta ^1-\zeta^0,
\partial_\sigma\zeta^1+\frac{1}{c}\zeta^4 \right)\,, \qquad 
\nonumber\\&&
D_i\zeta^4=\left(\partial_\tau\zeta^4+\frac{1}{s}\zeta^0, 
\partial_\sigma\zeta^4-\frac{1}{c}\zeta^1 \right)\,, \qquad 
D_i\zeta^{a}=\partial_i\zeta^{a} \qquad 
(a=2,3,5,\cdots,9)\,, \nonumber \\ 
&& X^{AB}=\diag\left(\frac{1}{s^2},\frac{2}{s^2}-\frac{1}{c^2},
\frac{2}{s^2},\frac{2}{s^2},\frac{2}{s^2}-\frac{1}{c^2s^2}; 
0,0,0,0,0\right)
+\frac{2}{c^2s}\delta^{(A}_1\delta^{B)}_4\,. \nonumber 
\end{eqnarray}
Here the following abbreviations have been introduced: 
\begin{eqnarray}
 s \equiv \sinh\sigma\,, \qquad c \equiv \cosh\sigma\,. 
\label{abb}
\end{eqnarray}

\medskip 

Then the mass term for $\zeta^1$ and $\zeta^4$ can be expressed as 
\begin{eqnarray*}
(\zeta^1,\zeta^4)\left(
  \begin{array}{cc}
    \frac{2}{s^2}-\frac{1}{c^2} &\frac{1}{c^2s}    \\
    \frac{1}{c^2s}   &\frac{2}{s^2}-\frac{1}{c^2s^2}    \\
  \end{array}
\right)
\left(
  \begin{array}{c}
    \zeta^1   \\
    \zeta^4   \\
  \end{array}
\right)\,,
\end{eqnarray*}
and it is diagonalizable 
by taking the following linear combination
\begin{eqnarray}
\left(
  \begin{array}{c}
    \tilde\zeta^1   \\
    \tilde\zeta^4   \\
  \end{array}
\right)=
\left(
  \begin{array}{cc}
   \frac{s}{c}    &-\frac{1}{c}    \\
   \frac{1}{c}    &  \frac{s}{c}  \\
  \end{array}
\right)
\left(
  \begin{array}{c}
    \zeta^1   \\
    \zeta^4   \\
  \end{array}
\right)\,. \nonumber 
\end{eqnarray}
The mass eigenvalues are given by 
\begin{eqnarray}
\tilde X^{AB}=\diag\left
(\frac{1}{s^2},\frac{1}{s^2},\frac{2}{s^2},\frac{2}{s^2},
\frac{2}{s^2}; 0,0,0,0,0\right)\,. \nonumber 
\end{eqnarray}
The new linear combination has been introduced and then 
the covariant derivative accordingly turn to be 
\begin{eqnarray}
&&
D\zeta^0=\nabla\zeta^0~, \qquad 
D\zeta^1=\left(\frac{s}{c}\nabla_\tau\tilde\zeta^1+\frac{1}{c}
\partial_\tau\tilde\zeta^4,
\frac{s}{c}\nabla_\sigma\tilde\zeta^1
+\frac{1}{c}\partial_\sigma\tilde\zeta^4\right)~,
\nonumber\\&&
D\zeta^4=\left(
-\frac{1}{c}\nabla_\tau\tilde\zeta^1+\frac{s}{c}\partial_\tau\tilde\zeta^4,
-\frac{1}{c}\nabla_\sigma\tilde\zeta^1+\frac{s}{c}
\partial_\sigma\tilde\zeta^4 \right)~, \nonumber 
\end{eqnarray}
where we have defined
\begin{eqnarray}
\nabla_i\zeta^0
\equiv 
\left(
\partial_\tau\zeta^0-\frac{c}{s}\tilde\zeta^1,\partial_\sigma\zeta^0
\right)\,, 
\qquad
\nabla_i\tilde\zeta^1 \equiv 
\left(\partial_\tau\tilde\zeta^1+\frac{c}{s}\zeta^0,
\partial_\sigma\tilde\zeta^1\right)\,. \nonumber 
\end{eqnarray}
Substituting the above quantities into the action, the resulting action is 
\begin{eqnarray}
S_{\rm 2B}&=&\frac{1}{4\pi}\int\! d^2\xi\,\Big[
\delta^{ij}D_i \hat \zeta^AD_j \hat \zeta^A
+\tilde X_{AB}\hat\zeta^A\hat \zeta^B
\Big]~, \nonumber  \\ 
&& \hat\zeta^A=(\zeta^0,\tilde\zeta^1,\zeta^2,\zeta^3,
\tilde\zeta^4,\zeta^5,\cdots,\zeta^9)\,, 
\quad D\tilde\zeta^{1}=\nabla\tilde\zeta^{1}\,, \quad 
D\tilde\zeta^{4}=\partial\tilde\zeta^{4}\,. \nonumber 
\end{eqnarray}
Let us write it as the action on the two-dimensional induced metric
\begin{eqnarray}
g_{ij}=\frac{1}{s^2}\delta_{ij}\,, \qquad R^{(2)}=-2\,,
\label{induced metric 1}
\end{eqnarray}
so that 
\begin{eqnarray}
S_{\rm 2B}
&=&
\frac{1}{4\pi}\int \!d^2\xi \sqrt{g}\Big[
g^{ij}D_i \hat \zeta^AD_j \hat \zeta^A
+s^2\tilde X_{AB}\hat\zeta^A\hat \zeta^B
\Big]~. \nonumber 
\end{eqnarray}

\medskip

Imposing Virasoro constraints is equivalent to removing the longitudinal
modes by adding the ghost action
\begin{eqnarray}
S_{\rm gh}&=&
\frac{1}{2}\int\! d^2\xi \sqrt{g}\left[
g^{ij}\nabla_i\epsilon^\alpha\nabla_j\epsilon^\alpha
-\frac{1}{2}R^{(2)}\epsilon^\alpha\epsilon^\alpha
\right]\,. 
\label{ghost action}
\end{eqnarray}
We choose $g_{ij}$ as the induced metric (\ref{induced metric 1}).
The covariant derivative is defined as 
\[
\nabla_i\epsilon^\alpha
\equiv \partial_i\epsilon^\alpha+\omega^\alpha_i{}_\beta\epsilon^\beta\,, 
\]
where $\omega$ is the two-dimensional spin connection:
$\omega^0{}_1=-\frac{c}{s}d\tau$, then
\begin{eqnarray}
\nabla_i\epsilon^0=\left(\partial_\tau\epsilon^0-\frac{c}{s}\epsilon^1,
\partial_\sigma\epsilon^0\right)\,,
 \qquad 
\nabla_i\epsilon^1=\left(\partial_\tau\epsilon^1+\frac{c}{s}\epsilon^0,
\partial_\sigma\epsilon^1 \right)~.
\nonumber 
\end{eqnarray}
The ghost action is the same as that of $\zeta^0$ and
$\tilde\zeta^1$\,. Thus these modes may be eliminated by the
constraints. The final gauge-fixed action is
\begin{eqnarray}
&& S_{\rm 2B} = \frac{1}{4\pi}\int \!d^2\xi \sqrt{g}\Big[
g^{ij}\partial_i \hat \zeta^A\partial_j \hat \zeta^A
+2(\zeta^2)^2+2(\zeta^3)^2+2(\tilde \zeta^4)^2
\Big]~, \nonumber \\
&& \qquad\qquad \quad 
\hat\zeta^A=(\zeta^2,\zeta^3,\tilde\zeta^4,\zeta^5,\dots,\zeta^9)\,.
\label{result}
\end{eqnarray}
This action contains three massive bosons with $m^2=2$ and five massless
bosons propagating on E\ads{2}. The fluctuations respect an $SO(3)\times
SO(5)$ symmetry, which is also preserved by the circular Wilson loop.

\medskip 

The resulting action (\ref{result}) is the same as the straight Wilson
line case. This result is not so surprising since the difference between
a straight Wilson line and a circular Wilson loop is the behavior at
infinity and it only gives an anomalous contribution to the value of the
action. However the local behavior around a finite point should not be
different.

\medskip

Finally, let us comment on the fermionic fluctuations. The quadratic
action is
\begin{eqnarray}
&& S_{\rm 2F}=\frac{i}{2\pi}\int\! d^2\xi 
\left[\sqrt{g}g^{ij}\delta^{IJ}-\epsilon^{ij}\sigma_3^{IJ}\right]
\bar\theta^I\rho_iD_j\theta^J\,, \nonumber \\ 
&& D_i\theta^I=\partial_i\theta^I+\frac{1}{4}\Omega_i^{AB}\Gamma_{AB}\theta^I
-\frac{i}{2}\epsilon^{IJ}(E_i^a\Gamma_a+iE_i^{a'}\Gamma_{a'})\theta^J \,, 
\qquad \rho_i=E_i^A\Gamma_A\,, \nonumber 
\end{eqnarray}
where $E$ and $\Omega$ are evaluated with the classical solution. After
rotating the spinor basis so that a two-dimensional spinor covariant
derivative is manifest, and fixing $\kappa$-symmetry appropriately,
the mass squared for the fermions is $m^2=1$ and the mass term is
proportional to $\bar\vartheta\Gamma_{01}\vartheta$ \cite{DGT} (where
$\vartheta$ is the rotated spinor). Thus the $SO(3)\times SO(5)$
symmetry is not broken by the fermions.

\section{Small deformations of circular Wilson loop}

From now on let us discuss the corresponding gauge-theory operator 
describing the fluctuations obtained in the previous section by
following \cite{Miwa, SY:non-rela2}. 

\medskip 

Let us consider a Wilson loop
\begin{eqnarray}
W(C)&=&\Tr P \CW\,, \qquad 
\CW \equiv \exp\oint\!\!ds (i A_\mu\dot x^\mu+\phi_i\dot y^i )
= \exp  \oint \!\!ds (i A_M\dot Y^M )~,
\label{WL}
\end{eqnarray}
where $\dot Y^M=(\dot x^\mu,-i \dot y^i)$\,. 
The supersymmetry transformation
\[
\delta_\epsilon A_\mu=i\bar\Psi\Gamma_\mu\epsilon\,, \qquad 
\delta_\epsilon\phi_i=i\bar\Psi \Gamma_i \epsilon\,, 
\] 
gives the following expression
\begin{eqnarray}
\delta_\epsilon W(C)&=&
\Tr P[i\delta_\epsilon A_M \dot Y^M \CW]
=\Tr P[ -\dot Y^M \bar\Psi\Gamma^M\epsilon \CW]~. \nonumber 
\end{eqnarray}
Thus the Wilson loop (\ref{WL}) is invariant under supersymmetry
transformation if
$\dot Y^M \Gamma_M\epsilon=0$.
The locally supersymmetry condition is derived
as the integrability
\begin{eqnarray}
(\dot Y^M \Gamma_M)^2=\dot Y^M\dot Y^N\eta_{MN}
=(\dot x^\mu)^2-(\dot y^i)^2=0\,. \nonumber 
\end{eqnarray}

We are interested in a circular Wilson loop configuration $C_0$
\begin{eqnarray}
x^\mu_{C_0}=(R\sin s,R(\cos\tau-1),0,0)\,, \qquad 
\dot y^i_{C_0}=(0,0,0,0,0,R) \nonumber 
\end{eqnarray}
where $R$ is the radius of the loop. Because
$A_0dx^0+A_1dx^1=A_rdr+A_\theta r d\theta $\,, we have
\begin{eqnarray*}
[iA_\mu\dot x^\mu+\phi_i\dot y^i]|_{C_0}
=R(iA_\theta+\phi_6)\,. \nonumber 
\end{eqnarray*}
Hence the Wilson loop $W(C_0)$ is
\begin{eqnarray}
W(C_0)=\Tr P[\CW_{C_0}]\,, \qquad 
\CW_{C_0}=
\exp\oint ds R (iA_\theta+\phi_6)\,, \nonumber 
\end{eqnarray}
which satisfies the locally supersymmetric condition. We identify
$W(C_0)$ as the vacuum operator.

\bigskip 

Let us consider a small deformation of $C_0$: $C=C_0+\delta C$
\begin{eqnarray}
x^\mu(C)=x^\mu(C_0)+ \delta x^\mu\,, \qquad 
\dot y^i(C)=\dot y^i(C_0)+\delta \dot y^i\,. \nonumber 
\end{eqnarray}
The Wilson loop can be expanded as
\begin{eqnarray}
W(C)&=&W(C_0)
+\oint ds \delta x^\mu(s)\frac{\delta W(C)}{\delta x^\mu(s)}\Big|_{C_0}
+\oint ds \delta \dot y^i(s)\frac{\delta W(C)}{\delta \dot y^i(s)}\Big|_{C_0}
+\cdots \nonumber\\
&=& W(C_0) +\oint ds\left(\delta x^\mu
\Tr P[i(F_{\mu N}\dot Y^{N})\CW]|_{C_0}
+\delta \dot y^i
\Tr P[\phi_i\CW]|_{C_0}
\right)
+\cdots\,, \nonumber 
\end{eqnarray}
where ellipsis implies higher order fluctuations. Note that
\begin{eqnarray}
\delta x^\mu(s)
\frac{\delta W(C)}{\delta x^\mu}\Big|_{C_0}&=&
\delta \theta(s)\Tr P\, R (D_\theta\phi_6)_s \CW_{C_0}
+\delta x^a(s)\Tr P\,(iF_{a\theta}+D_a\phi_6)_s\CW_{C_0}~, 
\label{delta W r}\\
\delta \dot y^{i}
\frac{\delta W(C)}{\delta \dot y^{i}}\Big|_{C_0}&=&
\delta \dot y^i 
\Tr P\, (\phi_i)_s \CW_{C_0}\,, 
\label{delta W y}
\end{eqnarray}
where $a=r,2,3$\,.

We require that a small deformation should satisfy the locally
supersymmetry condition, that is
\begin{eqnarray}
0 &=&
(\dot Y+\delta\dot  Y)^2=
(\dot x^\mu +\delta\dot x^\mu)^2
-(\dot y^i +\delta\dot y^i)^2\,. \nonumber 
\end{eqnarray}
This condition implies
\begin{eqnarray}
0=\dot x^\mu\delta\dot x^\mu-\dot y^i\delta\dot y^i
=R(\delta\dot\theta-\delta\dot y^6)
~. \nonumber 
\end{eqnarray}
On the other hand, by using an $SO(2)$\,, we can impose the condition to
the fluctuations,
\begin{eqnarray}
0=\delta\dot\theta+\delta\dot y^6\,,  \nonumber 
\end{eqnarray}
so that $\delta\dot\theta=0$ and $\delta\dot y^6=0$. The former means
$\delta\theta=0$. As a result we are left with impurities
\begin{eqnarray}
&&R(iF_{a\theta}+D_a\phi_6) \qquad (a=r,2,3)~, \qquad 
\phi_{a'} \qquad (a'=1,2,\cdots,5)~. 
\label{impurities NR}
\end{eqnarray}
Thus the resulting dictionary of impurity insertion is the same as in
the case of straight line, up to the appearance of the radius parameter
$R$.  This can be absorbed into the definition of $s$ by
$s\to\frac{s}{R}$ which corresponds to $\tau\to \frac{\tau}{R}$\,.

This impurity insertion respects an $SO(3)\times SO(5)$ symmetry, which
is also a symmetry of the quadratic action derived in section 2.  As
expected, the conformal dimensions of these impurities agree with the
mass dimensions of fluctuations $\Delta=\frac{1}{2}(1+\sqrt{1+4m^2})$:
\begin{eqnarray}
\Delta(\zeta^2,\zeta^3,\tilde\zeta^4)=2\,, \qquad 
\Delta(\zeta^i)=1 \quad (i=1,\cdots,5)\,. \nonumber 
\end{eqnarray}
The mass dimensions of fermionic fluctuations
$\Delta=\frac{1}{2}(1+2|m|)$ is
\begin{eqnarray*}
\Delta(\vartheta^\alpha)=\frac{3}{2}~~~ (\alpha=1,\cdots,8)\,, \nonumber 
\end{eqnarray*}
since $m^2=1$\,. Thus we expect that the eight fermionic impurities
with conformal dimension $\frac{3}{2}$ are inserted in the Wilson loop
as well as bosonic impurities (\ref{impurities NR}). This expectation is
correct as we show in Appendix B.

\section{Semiclassical limit around Miwa-Yoneya solution}

As the second issue we consider a rotating classical solution in which
the boundary is a circular Wilson loop with local operator
insertions. The classical solution was constructed in \cite{MY}. This
solution is a generalization of \cite{DK}, which was constructed for a
straight Wilson line in Lorentzian signature.  We will examine a
correspondence between the fluctuation about the classical solution and
small deformation of the circular Wilson loop with local operator
insertions.

\subsection{Classical solution for a circular Wilson loop with local 
operator insertions}
First of all, let see the classical solution found in \cite{MY}. We work
in Euclidean \adss{5}{5} with the Poincare coordinates (See appendix A
for vielbeins and spin connections.)
\begin{eqnarray}
ds^2=\frac{1}{z^2}(
dx_i^2
+dz^2)
+\cos^2\theta d\psi^2
+d\theta^2
+\sin^2\theta d\Omega_3^2~.\label{Pm}
\end{eqnarray} 

\medskip 

The solution corresponding to a circular Wilson loop with the local
operator insertions is given by \cite{MY}
\begin{eqnarray}
z=\frac{\ell\sinh\sigma}{\cosh\sigma\cosh\tau\pm\alpha}~, \quad 
x_0=\frac{\ell\cosh\sigma\sinh\tau}{\cosh\sigma\cosh\tau\pm\alpha}~, \quad 
x_1=\frac{\pm\ell\sqrt{1-\alpha^2}}{\cosh\sigma\cosh\tau\pm\alpha} \quad 
\label{classical sol MY1}
\end{eqnarray}
for the AdS$_5$ part, and 
\begin{eqnarray}
\psi = \tau\,, \qquad \cos\theta=\tanh\sigma
\label{classical sol MY2}
\end{eqnarray}
for the S$^5$ part. The parameter $\alpha$ parametrizes the radius of the
loop $R$ like 
\begin{eqnarray}
 R = \frac{\ell}{\sqrt{1-\alpha^2}}\,. \label{radius}
\end{eqnarray}
Note that we have performed a Wick rotation as $\psi \to -i\psi$
following \cite{tunneling,MY}. That is why we can define a sensible
angular momentum even in Euclidean signature. But we should keep it in
mind that the signature of $d\psi^2$ in the metric (\ref{Pm}) is flipped
as $-d\psi^2$ due to the Wick rotation of $\psi$\,.

We can easily see that (\ref{classical sol MY1}) and (\ref{classical sol
MY2}) solve the equation of motion (\ref{EL}), and the Virasoro
constraints (\ref{Vir}). The classical solution for the AdS$_5$ part
(\ref{classical sol MY1}) consists of the two patches corresponding to
the upper and lower signs in (\ref{classical sol MY1}).
The parameter $\alpha$ takes the value in  $0\le\alpha\le 1$\,.
When $\alpha \neq 1$\,, it is a circular Wilson loop,
while when $\alpha= 1$, a straight Wilson loop. For $\alpha=1$\,, 
we can see that $R=\infty$ from (\ref{radius})\,. 

The boundary of the string worldsheet is at $z=0$. 
The Wilson loop at $z=0$ corresponds to
 $(\sigma,\tau)=(0,\tau)$ and $(\sigma,\tau)=(\infty, \pm\infty)$,
where local operators are inserted at latter two points.

\subsection{Semiclassical limit} 

Next lets us consider a semiclassical action around the solution
(\ref{classical sol MY1}) and (\ref{classical sol MY2})\,. 

\medskip 

The quadratic action around the classical solution (\ref{classical sol
MY1}) and (\ref{classical sol MY2}) is basically given by (\ref{fluc
2})\,, where the covariant derivatives and mass matrices are replaced by
\begin{eqnarray}
&&
D_i\zeta^0=\left(\partial_\tau\zeta^0-\frac{\dot x_0}{z}\zeta^4,
\partial_\sigma\zeta^0-\frac{x_0'}{z}\zeta^4 \right)~, \qquad 
D_i\zeta^1=\left(\partial_\tau\zeta^1-\frac{\dot x_1}{z}\zeta^4,
\partial_\sigma\zeta^1-\frac{x_1'}{z}\zeta^4\right)~,~~
\nonumber\\&&
D_i\zeta^4=\left(\partial_\tau\zeta^4+\frac{\dot x_0}{z}\zeta^0
+\frac{\dot x_1}{z}\zeta^1,
\partial_\sigma \zeta^4+\frac{x_0'}{z}\zeta^0
+\frac{x_1'}{z}\zeta^1\right)~,~~
\nonumber\\&&
D_i\zeta^5=\left(\partial_\tau\zeta^5-\sin\theta\dot\psi\zeta^6,
\partial_\sigma\zeta^5\right)~, \qquad 
D_i\zeta^6=\left(\partial_\tau\zeta^6+\sin\theta\dot\psi\zeta^5,
\partial_\sigma\zeta^6\right)~,~~
\nonumber\\&&
D_i\zeta^{a}=\partial_i\zeta^{a} \qquad (a=2,3,7,8,9)~, \nonumber 
\end{eqnarray}
and
\begin{eqnarray}
X^{ab}&=&
\diag\Big(
\frac{(\partial x_1)^2+(\partial z)^2}{z^2},
\frac{(\partial x_0)^2+(\partial z)^2}{z^2},
\frac{(\partial x_0)^2+(\partial x_1)^2+(\partial z)^2}{z^2},
\nonumber\\&&~~~~~~~
\frac{(\partial x_0)^2+(\partial x_1)^2+(\partial z)^2}{z^2},
\frac{(\partial x_0)^2+(\partial x_1)^2}{z^2}\Big)
\nonumber\\&&
-\frac{2}{z^2}\partial x_0\partial x_1\delta^{(A}_0\delta^{B)}_1
-\frac{2}{z^2}\partial x_0\partial z\delta^{(A}_0\delta^{B)}_4
-\frac{2}{z^2}\partial x_1\partial z\delta^{(A}_1\delta^{B)}_4~, \nonumber 
\\
X^{a'b'}&=&
\diag\Big(
-\theta'{}^2,
-\cos^2\theta\dot\psi^2,
-\cos^2\theta\dot\psi^2-\theta'{}^2,
-\cos^2\theta\dot\psi^2-\theta'{}^2,
-\cos^2\theta\dot\psi^2-\theta'{}^2
\Big)\,, \nonumber 
\end{eqnarray}
respectively. 
Here we should note that an additional term should be added to the action 
because of the presence of a conserved charge associated with $\psi$\,. 
The only effect of adding the term is to change the sign of the kinetic
term of $\psi$\,, and that is why we arrive at the same action even
after the Wick rotation of $\psi$\,.

\medskip 

Then we shall diagonalize the mass matrix for $\zeta^0$, $\zeta^1$ and
$\zeta^4$
\begin{eqnarray}
X&=&
\left(
  \begin{array}{ccc}
\frac{(\partial x_1)^2+(\partial z)^2}{z^2}  
&-\frac{\partial x_0\partial x_1}{z^2}    & 
-\frac{\partial x_0\partial z}{z^2}    \\
 -\frac{\partial x_0\partial x_1}{z^2}       
&\frac{(\partial x_0)^2+(\partial z)^2}{z^2}     & 
       -\frac{\partial x_1\partial z}{z^2}    \\
-\frac{\partial x_0\partial z}{z^2}         
&  -\frac{\partial x_1\partial z}{z^2}    
&\frac{(\partial x_0)^2+(\partial x_1)^2}{z^2}     \\
  \end{array}
\right)~. \nonumber 
\end{eqnarray}
For this we need to know the eigenvalues and eigenvectors
\begin{eqnarray}
Xv_{(i)}=e_{(i)}v_{(i)} \qquad (i=0,1,4)\,. \nonumber 
\end{eqnarray}
It is straightforward to derive the eigenvalues
\begin{eqnarray}
e_{(0)}=\frac{1}{\sinh^2\sigma}~, \quad 
e_{(1)}=\frac{\cosh^2\sigma}{\sinh^2\sigma}~, \quad 
e_{(4)}=\frac{1+\cosh^2\sigma}{\sinh^2\sigma}~. \nonumber 
\end{eqnarray}
The corresponding eigenvectors are found to be
\begin{eqnarray}
v_{(0)}^T&=& \left(
\frac{\alpha\cosh\tau+\cosh\sigma}{a},~
-\frac{\sqrt{1-\alpha^2}\sinh\tau}{a},~
-\frac{\sinh\tau\sin\sigma}{a} \right)~,~ \nonumber \\
v_{(1)}^T&=& \left(
\frac{\alpha\sinh\sigma\sinh\tau}{a},~
-\frac{\sqrt{1-\alpha^2}\sinh\sigma\cosh\tau}{a},~
\frac{\alpha\cosh\sigma+\cosh\tau}{a} \right)~,~  \nonumber \\
v_{(4)}^T&=& \left(
\frac{\sqrt{1-\alpha^2} \cosh\sigma\sinh\tau}{a},~
\frac{\alpha\cosh\sigma\sinh\tau+1}{a},~
\frac{\sqrt{1-\alpha^2} \sinh\sigma}{a} \right)~,~~~~\nonumber 
\end{eqnarray}
where 
\[
v_{(i)}^Tv_{(j)}=\delta_{ij}\,, \qquad a=\cosh\sigma\cosh\tau+\alpha\,. 
\]
Therefore we find
\begin{eqnarray}
V^{-1}XV= \diag(e_{(0)},e_{(1)},e_{(4)})\,, \qquad 
V=(v_{(0)},v_{(1)},v_{(4)})\,. \nonumber 
\end{eqnarray}
It follows that
\begin{eqnarray}
\zeta^T X\zeta = \tilde\zeta^T V^{-1}XV \tilde\zeta
=e_{(0)}(\tilde\zeta^0)^2
+e_{(1)}(\tilde\zeta^1)^2
+e_{(4)}(\tilde\zeta^4)^2\,,  \qquad 
\tilde\zeta= V^T\zeta\,. \nonumber 
\end{eqnarray}
Thus we have obtained the diagonalized mass-matrix
\begin{eqnarray}
\tilde X^{AB}=\diag\left(
\frac{1}{s^2}, \frac{c^2}{s^2}, \frac{c^2+1}{s^2},
\frac{c^2+1}{s^2}, \frac{c^2+1}{s^2}; \frac{-1}{c^2}, \frac{s^2}{c^2},
\frac{s^2-1}{c^2}, \frac{s^2-1}{c^2}, \frac{s^2-1}{c^2} 
\right)~. \nonumber
\end{eqnarray}
Here we have used the notation (\ref{abb})\,. 

\medskip 

Next let us examine the kinetic terms for $\zeta^0$, $\zeta^1$ and
$\zeta^4$\,. Then we see that
\begin{eqnarray}
D\zeta^0&=&
\frac{1}{a}\Big(
(\alpha\cosh\tau+\cosh\sigma)\nabla\tilde\zeta^0
+\alpha\sinh\sigma\sinh\tau\nabla\tilde\zeta^1
\nonumber\\&&~~~
+\sqrt{1-\alpha^2}\cosh\sigma\sinh\tau\partial\tilde\zeta^4
\Big)~,~~ \nonumber \\
D\zeta^1&=&
\frac{1}{a}\Big(
-\sqrt{1-\alpha^2}\sinh\tau\nabla\tilde\zeta^0
-\sqrt{1-\alpha^2}\sinh\sigma\cosh\tau\nabla\tilde\zeta^1
\nonumber\\&&~~~
+(\alpha\cosh\sigma\cosh\tau+1)\partial\tilde\zeta^4
\Big)~,~~\nonumber \\
D\zeta^4&=&
\frac{1}{a}\Big(
-\sinh\sigma\sinh\tau\nabla\tilde\zeta^0
+(\alpha\cosh\sigma+\cosh\tau)\nabla\tilde\zeta^1
 \nonumber\\&&~~~ 
+\sqrt{1-\alpha^2}\sinh\sigma\partial\tilde\zeta^4
\Big)\,, \nonumber 
\end{eqnarray}
where the following quantities have been introduced
\begin{eqnarray}
\nabla\tilde\zeta^0&\equiv& \left(\partial_\tau\tilde\zeta^0
-\frac{1}{\sinh\sigma}\tilde\zeta^1,\partial_\sigma\tilde\zeta^0
\right)~, 
\qquad 
\nabla\tilde\zeta^1\equiv \left(\partial_\tau\tilde\zeta^1
+\frac{1}{\sinh\sigma}\tilde\zeta^0,\partial_\sigma\tilde\zeta^1
\right)~.
\label{2-dim cd}
\end{eqnarray}
These expressions lead us to 
\begin{eqnarray}
(D\zeta^0)^2+(D\zeta^1)^2+(D\zeta^4)^2
=(\nabla\tilde\zeta^0)^2+(\nabla\tilde\zeta^1)^2
+(\partial\tilde\zeta^4)^2
~. \nonumber 
\end{eqnarray}
Thus we have obtained
\begin{eqnarray}
S_{\rm 2B}&=&\frac{1}{4\pi}\int\! d^2\xi\,[
D\tilde \zeta^A D\tilde\zeta^A+\tilde X^{AB}\tilde\zeta^A\tilde\zeta^B
]~, \qquad D\tilde\zeta^{0,1}=\nabla\tilde\zeta^{0,1}\,, \quad 
D\tilde\zeta^4=\partial \tilde\zeta^4\,, 
\nonumber \\ 
&& \qquad  \tilde \zeta^A=(\tilde\zeta^0, \tilde\zeta^1, \zeta^2, \zeta^3,
\tilde\zeta^4;\zeta^5,\cdots,\zeta^9)\,. 
\nonumber 
\end{eqnarray}

\bigskip 

We have derived the eigenvalues of the mass matrix in the bosonic sector
so far. It seems to be difficult to add the ghost action so that it
should delete the unphysical longitudinal modes in the matrix. Actually,
we have not completed this step and we will leave it as a future
problem. Instead of trying to delete the unphysical modes from the full
action, let us consider the Lagrangian density in the two special
regions, 1) $\sigma=0$ and 2) $\sigma=\infty$\,.  
For the former case we expect
that the Lagrangian density should behave as that in section 2. For the
latter case the Lagrangian density is expected to be the one for the
pp-wave string action.

\subsection*{The fluctuations near $\sigma=0$}

The ghost action (\ref{ghost action}) is introduced as before. 
Let us consider the two-dimensional metric 
\begin{eqnarray}
g_{ij}=\frac{\cosh^2\sigma}{\sinh^2\sigma}d\tau^2
+\frac{1}{\sinh^2\sigma}d\sigma^2\,, \quad R^{(2)}=-2\,, 
\label{2-dim metric}
\end{eqnarray}
which is \ads{2} and corresponds to the induced metric of \ads{5} part only.
The covariant derivative is defined by
\[
\nabla_i\epsilon^\alpha \equiv 
\partial_i\epsilon^\alpha+\omega^\alpha_i{}_\beta\epsilon^\beta\,, 
\]
where $\omega$ is the two-dimensional spin connection:
$\omega^0{}_1=-\frac{1}{s}d\tau$, then
\begin{eqnarray}
\nabla_i\epsilon^0
=\left(\partial_\tau\epsilon^0-\frac{1}{\sinh\sigma}\epsilon^1,
 \partial_\sigma\epsilon^0 \right)~, \qquad 
\nabla_i\epsilon^1
= \left(\partial_\tau\epsilon^1+\frac{1}{\sinh\sigma}\epsilon^0,
\partial_\sigma\epsilon^1 \right)~. \nonumber 
\end{eqnarray}
Thus we see that the covariant derivative defined in (\ref{2-dim cd})
is nothing but two-dimensional covariant derivative.

The mass dimension of a fluctuation is derived through its 
behavior near the boundary.  Hence we examine fluctuations near
$\sigma=0$\,. Let us rewrite $S_{\rm 2B}=\int d^2\xi\CL_{\rm 2B}$ by
using (\ref{2-dim metric}). Near $\sigma=0$
\begin{eqnarray}
\CL_{\rm 2B}&\approx &\frac{1}{4\pi}
\sqrt{g}[
g^{ij}D_i\tilde \zeta^A D_j\tilde\zeta^A
+(\tilde\zeta^0)^2
+(\tilde\zeta^1)^2
+2\sum_{i=2,3,4}(\tilde\zeta^i)^2
]~~. \nonumber 
\end{eqnarray}
Here we note that the contribution of fluctuations $\tilde\zeta^0$ and
$\tilde\zeta^1$ is the same form as the ghost action near $\sigma=0$, so
that it is canceled out by the ghost contribution.  In addition we
should note that $g^{ij}D_i\tilde\zeta^aD_j\tilde\zeta^a\approx
g^{ij}\partial_i\tilde\zeta^a\partial_j\tilde\zeta^a$ ($a=5,6$).

\medskip

As a result, we are left with three massive bosons with $m^2=2$ and five
massless bosons propagating in \ads{2}. The symmetry preserved by the
fluctuations is $SO(3)\times SO(5)$\,. The mass dimensions
$\Delta=\frac{1}{2}(1+\sqrt{1+4m^2})$ corresponding to the fluctuations
are
\begin{eqnarray}
\Delta(\tilde \zeta^i)=2 \quad (i=2,3,4)~, \qquad 
\Delta(\tilde\zeta^{a'})=1 \quad (a'=5,6,\cdots,9)~. \nonumber 
\end{eqnarray}
Thus we have reproduced the dictionary obtained in section 3 and
\cite{SY:non-rela} as expected. As the induced metric reduces to
(\ref{induced metric 1}) near $\sigma=0$, the fermionic fluctuations are
the same as those given in section 2.2. The mass dimensions of fermionic
fluctuations are $\Delta(\vartheta^\alpha)=\frac{3}{2}$
($\alpha=1,\cdots,8$)\,.

\subsection*{The fluctuations near $\sigma=\infty$}

In the region near $\sigma=\infty$ we should introduce the ghost action
(\ref{ghost action}) with flat two-dimensional metric
\begin{eqnarray}
g_{ij}=d\tau^2
+d\sigma^2~, \qquad 
R^{(2)}=0~.
\label{2-dim metric at sigma=infty}
\end{eqnarray}
Then we can see that the longitudinal modes are canceled out with the 
ghost action as follows.

First let us consider the fluctuations in the neighbor of
$\sigma=\infty$\,. Then by using (\ref{2-dim metric at sigma=infty})
$\CL_{\rm 2B}$ can be rewritten as 
\begin{eqnarray}
\CL_{\rm 2B}&\approx &\frac{1}{4\pi}
\sqrt{g}[
g^{ij}D_i\tilde \zeta^A D_j\tilde\zeta^A
+\sum_{i\neq 0,5
}(\tilde\zeta^i)^2
]~~, \nonumber 
\end{eqnarray}
where $D_i\tilde\zeta\approx \partial_i\tilde\zeta$\,. 
The contribution of fluctuations $\tilde\zeta^0$ and $\tilde\zeta^5$ is
the same form as the ghost action, hence it is canceled out by the ghost
contribution. Thus we are left with eight massive bosons with $m^2=1$
propagating in the two-dimensional flat space. 
The fermionic fluctuations
with $m^2=1$ break $SO(8)$ to $SO(4)\times SO(4)$
(see for example \cite{FT}). 
This is nothing but 
the Lagrangian density of a pp-wave string.

\section{Small deformations of circular Wilson loop}

Let us consider the quadratic fluctuations discussed in the previous
section as a small deformation of a circular Wilson loop with local
operator insertions. 

\medskip 

The classical solution given by (\ref{classical sol MY1}) and
(\ref{classical sol MY2}) is attaching to the boundary. 
The boundary of the classical string worldsheet can be seen as a circle 
on $\mathbb{R}^4$\,. The classical solution on the boundary is described by 
\begin{eqnarray}
x^\mu=
\left(\frac{\ell\sinh\tau}{\cosh\tau\pm\alpha}, 
\frac{\pm\ell\sqrt{1-\alpha^2}}{\cosh\tau\pm\alpha},0,0
\right)~\,, \nonumber 
\end{eqnarray}
and the circle is represented by 
\begin{eqnarray}
(x_0)^2+(x_1+\alpha R)^2=R^2~, \qquad 
R=\frac{\ell}{\sqrt{1-\alpha^2}}\,. \nonumber 
\end{eqnarray}
That is, the radius of the loop is $R$ and its center is located at
$(x_0,x_1)=(0,-\alpha R)$\,. It is also convenient to introduce radial
coordinates $r$ and $\theta$
\begin{eqnarray}
\tilde x^\mu=
(x_0,x_1+\alpha R,x_2,x_3)
=(r\sin\theta,r\cos\theta,x_2,x_3)~. \nonumber 
\end{eqnarray}
Then the circle lies along $(r,\theta)=(R,s)$\,.

\medskip 

We are now considering the local operator insertions $Z^J$ and $\bar
Z^J$\,. Here we define $Z \equiv\phi_1+i\phi_2$ and $\bar Z
\equiv\phi_1-i\phi_2$\,. As explained in \cite{MY}, $Z^J$ and $\bar Z^J$
are inserted at $\tau=-\infty$ and $\tau=+\infty$, respectively. 
That is, $Z^J$ and $\bar Z^J$ are inserted at $s=s_1, s_2$ with 
$(x_0,x_1)=(-\ell,0)$ and $(+\ell,0)$\,, respectively.

\medskip 

Let us consider the Wilson loop
\begin{eqnarray}
W(C)&=&
\Tr \CW
\,, \quad 
\CW=P\,
\exp\oint ds \left(
iA_\mu(x(s)) \dot{\tilde  x}^\mu(s)
+\phi_i(x(s)) \dot y^i(s)
\right)\,.  \nonumber 
\end{eqnarray}
The following expressions are useful later:
\begin{eqnarray}
\frac{\delta}{\delta x^\mu(s)}\CW&=&
P
(iF_{\mu\nu}\dot x^\nu(s)+D_\mu\phi_i\dot y^i(s))
\CW
~, \qquad 
\frac{\delta}{\delta \dot y^i(s)}\CW=
P
\phi_i(s)\CW~. \nonumber 
\end{eqnarray} 
Let us take $C_0$ as
\begin{eqnarray}
C_0~:~~\left\{
  \begin{array}{ll}
\tilde x^\mu=(R\sin s, R\cos s,0,0)~,~
\dot y^i=(0,0,0,0,0,R)        & s\neq s_1,s_2    \\
\tilde x^\mu=(R\sin s_1, R\cos s_1,0,0)~,~
\dot y^i=(1,i,0,0,0,0)        & s= s_1    \\
\tilde x^\mu=(R\sin s_2, R\cos s_2,0,0)~,~
\dot y^i=(1,-i,0,0,0,0)        ~~~& s= s_2    \\
  \end{array}
\right.
~, \nonumber 
\end{eqnarray}
then
\begin{eqnarray}
W(C_0)&=&
\Tr P\left[
\e^{\oint Z}(s_1)
\e^{\oint \bar Z}(s_2)
\exp\oint ds ~R(iA_\theta(\theta(s))+\phi_6(\theta(s)) )\right]~
\cr&=&
\sum_{J,J'}\frac{1}{J!J'!}
\Tr P\left[
 Z(s_1)^J
\bar Z(s_2)^{J'}
\exp\oint ds ~R(iA_\theta(\theta(s))+\phi_6(\theta(s)) )\right]\,, 
\nonumber 
\end{eqnarray}
which does not vanish only when $J=J'$\,. 

\medskip 

\begin{figure}[htbp]
 \begin{center}
  \includegraphics[scale=.7]{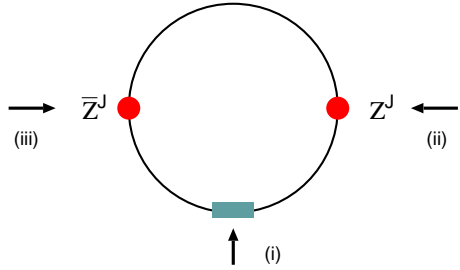}
\caption{\footnotesize The impurity insertions onto the circular Wilson
  loop with the local operator insertions. Three types of insertions can
  be considered. For (i) the insertion obeys the rule for the circular
  case without the local operators. For (ii) the insertion rule is given
  by the BMN dictionary. The case (iii) is a complex conjugate of (ii).}
\label{Wilson:fig}
 \end{center}
\end{figure}

We may consider three kinds of impurity insertions as depicted 
in Fig.\,\ref{Wilson:fig}. Let us consider each of the cases below.

\bigskip 

\subsubsection*{A small deformation at $s\neq s_1,s_2$}

Let us consider a small deformation of $C_0$ at $s\neq s_1,s_2$. This
corresponds to the case (i) in Fig.\,\ref{Wilson:fig}. 
The
Wilson loop $W(C)$ is expanded as
\begin{eqnarray}
W(C)&=&
W(C_0)
+
\oint ds\left(\delta x^\mu
\Tr P[(iF_{\mu \nu}\dot x^\nu+D_\mu\phi_i\dot y^i)\CW]|_{C_0}
+\delta \dot y^i
\Tr P[\phi_i\CW]|_{C_0}
\right)
+\cdots \nonumber 
\end{eqnarray}
where ellipsis are higher order fluctuations.
One derives equations (\ref{delta W r}) and (\ref{delta W y})
with 
\begin{eqnarray}
\CW_{C_0}&=&
\sum_{J}\frac{1}{J!J!}
Z(s_1)^J
\bar Z(s_2)^J
\exp\oint ds ~
R(iA_\theta+\phi_6)
~. \nonumber 
\end{eqnarray}

As in section 3, by requiring that the small deformation should meet
locally supersymmetry condition $0= (\dot Y+\delta\dot Y)^2 $, and by
using an $SO(2)$, we have $\delta\dot y^6=0$ and $\delta\theta=0$.  As a
result we are left with impurities (\ref{impurities NR})~.  As expected,
the conformal dimensions of these impurities agree with the mass
dimensions of fluctuations $\Delta=\frac{1}{2}(1+\sqrt{1+4m^2})$:
\begin{eqnarray}
\Delta(\tilde \zeta^i)=2~~~(i=2,3,4)~, \qquad 
\Delta(\tilde\zeta^{a'})=1~~(a'=5,\cdots,9)\,. \nonumber 
\end{eqnarray}
This is nothing but the dictionary for the gauge-theory operators 
corresponding to the fluctuations around the circular solution 
(non-relativistic string) \cite{SY:non-rela2}.

\subsubsection*{A small deformation at $s= s_1,s_2$}

Next let us expand $W(C)$ around $C_0$ at $s=s_1$\,.  This
corresponds to the cases (ii) and (iii) in Fig.\,\ref{Wilson:fig}. 
The argument here basically follows \cite{Miwa}. 

\medskip 

In this case the Wilson loop can be expanded as
\begin{eqnarray}
W(C)&=&W(C_0)
+\oint ds
\Bigl( \delta x^\mu \Tr P[ D_\mu Z(s_1) \CW]|_{C_0}
+\delta\dot y^i \Tr P[ \phi_i(s_1) \CW]|_{C_0}
\Bigr) ~+~\cdots\,. \nonumber 
\end{eqnarray} 
By requiring that the small fluctuations should satisfy the locally
supersymmetry condition, we obtain the following condition,
\begin{eqnarray}
\delta \dot y^1 +i \delta \dot y^2=0\,. \nonumber 
\end{eqnarray}
In addition a reparametrization invariance allows us to fix the
fluctuations as
\[
\delta \dot y^1 -i \delta \dot y^2=0\,.
\]
Thus we can impose the condition $\delta \dot y^1= \delta \dot y^2=0$\,.

As a result, we are left with impurities inserted at $s=s_1$
\begin{eqnarray}
&&D_\mu Z~~~(\mu=0,1,2,3)~, \qquad 
\phi_i~~~(i=3,4,5,6)~. \nonumber 
\end{eqnarray}
Similarly, for a small deformation at $s=s_2$ we have the impurities
inserted at $s=s_2$
\begin{eqnarray}
&&D_\mu \bar Z~~~(\mu=0,1,2,3)~, \qquad 
\phi_i~~~(i=3,4,5,6)~. \nonumber 
\end{eqnarray} 
The impurities respect $SO(4)\times SO(4)$ symmetry. 
This dictionary is nothing but the BMN one. 

\bigskip

In summary, the resulting dictionary says that 
the impurity insertion at the local operators should follow the
BMN-dictionary and other than the position of local operators it should
follow the dictionary we obtained in \cite{SY:non-rela2}.

\section{Summary and Discussion}

We have discussed a semiclassical approximation around the classical
solutions corresponding to circular Wilson loops without/with local
operator insertions $Z^{J}$ and $\bar{Z}^J$\,. We have derived a
quadratic action for each of the cases and identified the fluctuations
with small deformations of the Wilson loop. With the local insertions
the action behaves as the non-relativistic string around $\sigma=0$\,,
while it as the pp-wave string around $\sigma=\infty$\,.

\medskip 

The dictionary of operator insertion has been clarified from the
viewpoint of a small deformation of Wilson loops without/out local
operator insertions. Without the local operators it is the same as in
the case of straight Wilson line. With them we have discussed it around
$\sigma=0$ and $\sigma=\infty$\,. Near $\sigma=0$ it is the same as in
the case of the circular Wilson loop without the local operators.  Near
$\sigma=\infty$ it is nothing but the BMN one as expected. This result
nicely agrees with the behavior of the quadratic fluctuations.

\medskip 

For the case of the circular Wilson loop with local operator insertions,
it remains as a problem to be solved to find a ghost terms to remove the
unphysical longitudinal modes for an arbitrary $\sigma$\,. It would be
interesting to compute a semiclassical partition function after solving
these problems.

\medskip 

One may consider a spin chain description for the circular Wilson loop
case. The gauge-theory analysis is discussed in \cite{DK}. The
remaining work is to construct circular Wilson loop solutions rotating
with two or more spins and compare them with the gauge-theory results.
A part of this issue has already been discussed in \cite{Tsuji}.

\medskip 

It is also nice to consider a semiclassical analysis to a dual giant
Wilson loop \cite{DF}. A rotating dual giant Wilson loop solution has
been already constructed in \cite{DGRT}. By following this paper, one
can discuss the semiclassical limit around the solution and it is
possible to deduce the corresponding gauge-theory side. 

\medskip 

We believe that our approach would give a new window to test the
AdS/CFT duality.

\subsection*{Acknowledgements}

The authors thank T.~Azeyanagi, S.~Iso, Y.~Kimura, Y.~Michishita,
A.~Miwa, K.~Murakami H.~Shimada, F.~Sugino, Y.~Sumitomo, R.~Suzuki, 
T.~Takayanagi, D.~Trancanelli
and S.~Yamaguchi for useful discussion. They also thank the Yukawa
Institute for Theoretical Physics at Kyoto University. Discussions
during the YITP workshop YITP-W-07-05 on ``String Theory and Quantum
Field Theory'' were useful to complete this work. They thank especially
A.~Miwa for valuable comments on the part of Wilson loop expansion.

The work of M.S.\ was supported in part by the Grant-in-Aid for
Scientific Research (19540324,\,19540304,\,19540098) from the Ministry
of Education, Science and Culture, Japan. The work of K.Y.\ was
supported in part by JSPS Postdoctoral Fellowships for Research Abroad
and the National Science Foundation under Grant No.\,NSF PHY05-51164.

\appendix

\section*{Appendix}

\section{Vielbeins and spin connections for AdS$_5\times$S$^5$}
Let us summarize vielbeins and spin connections of
AdS$_5\times$S$^5$\,. The metric utilized in section 2 is slightly
different from the one in section 4. In both of them we set the common
radius of AdS$_5$ and S$^5$ to be 1. 

\medskip 

\subsubsection*{The metric  in section 2} 
The E\ads{5} metric in the Poincare coordinate system, which was used in
section 2, is given by
\begin{eqnarray}
ds^2=\frac{1}{z^2}(
r^2d\theta^2
+dr^2
+dx_2^2
+dx_3^2
+dz^2)\,. \nonumber 
\end{eqnarray}
Here the two-dimensional subspace is described by a polar coordinate. 
The vielbein and spin connection are 
\begin{eqnarray}
E^a&=&\left(
\frac{r}{z}d\theta,
\frac{1}{z}dr,
\frac{1}{z}dx_2,
\frac{1}{z}dx_3,
\frac{1}{z }dz
\right)~,\nonumber\\
\Omega^1{}_4&=&
-\frac{1}{z}dr~,~~
\Omega^0{}_4=-\frac{r}{z}d\theta~,~~
\Omega^0{}_1=d\theta~,~~
\Omega^2{}_4=-\frac{1}{z}dx_2~,~~
\Omega^3{}_4=-\frac{1}{z}dx_3~. \nonumber 
\end{eqnarray}
Then we choose the metric  of \s{5}  as
\begin{eqnarray}
ds^2=d\varphi_1^2
+\cos^2\varphi_1(d\varphi_2^2+\cos^2\varphi_2(d\varphi_3^2+\cos^2\varphi_3(
d\varphi_4^2+\cos^2\varphi_4d\varphi_5^2)))\,. \nonumber 
\end{eqnarray}
The vielbein and spin connection are 
\begin{eqnarray}
E^{a'}&=&
(d\varphi_1,\,
\cos\varphi_1d\varphi_2,\,
\cos\varphi_1\cos\varphi_2d\varphi_3,\,
\nonumber\\&&~~~
\cos\varphi_1\cos\varphi_2\cos\varphi_3d\varphi_4,\,
\cos\varphi_1\cos\varphi_2\cos\varphi_3\cos\varphi_4d\varphi_5\,
)~,  \nonumber \\
\Omega^6{}_5&=&
-\sin\varphi_1d\varphi_2~,~~
\Omega^7{}_5=-\sin\varphi_1\cos\varphi_2d\varphi_3~,~~
\Omega^7{}_6=-\sin\varphi_2d\varphi_3~,~~
\nonumber\\
\Omega^8{}_5&=&-\sin\varphi_1\cos\varphi_2\cos\varphi_3d\varphi_4~,~~
\Omega^8{}_6=-\sin\varphi_2\cos\varphi_3d\varphi_4~,~~
\Omega^8{}_7=-\sin\varphi_3d\varphi_4~,
\nonumber\\
\Omega^9{}_5&=&-\sin\varphi_1\cos\varphi_2\cos\varphi_3\cos\varphi_4d\varphi_5~,~~
\Omega^9{}_6=-\sin\varphi_2\cos\varphi_3\cos\varphi_4d\varphi_5
\nonumber\\
\Omega^9{}_7&=&-\sin\varphi_3\cos\varphi_4d\varphi_5~,~~
\Omega^9{}_7=
-\sin\varphi_4d\varphi_5~. \nonumber
\end{eqnarray}

\medskip

\subsubsection*{The metric in section 4} 
The E\ads{5} metric in the Poincare coordinate system, which was used in
section 4, is given by
\begin{eqnarray}
ds^2=\frac{1}{z^2}(dx_i^2 +dz^2)\,,  \nonumber 
\end{eqnarray}
where $i=0,\cdots,3$ and the four-dimensional subspace is described by 
Cartesian coordinates. The vielbein and spin connection are 
\begin{eqnarray}
E^a&=&\left(
\frac{1}{z}dx_0,
\frac{1}{z}dx_1,
\frac{1}{z}dx_2,
\frac{1}{z}dx_3,
\frac{1}{z }dz
\right)~,\nonumber\\
\Omega^i{}_4&=&
-\frac{1}{z}dx_i~~(i=0,\cdots,3)~. \nonumber 
\end{eqnarray}
Then the metric  of \s{5} is chosen as 
\begin{eqnarray}
ds^2=\cos^2\theta d\psi^2
+d\theta^2
+\sin^2\theta
(d\varphi_1^2+\cos^2\varphi_1(d\varphi_2^2+\cos^2\varphi_2d\varphi_3^2))
~. \nonumber 
\end{eqnarray}
The vielbein and spin connection are computed as follows:  
\begin{eqnarray}
E^{a'}&=&
(\cos\theta d\psi,\,
d\theta,\,
\sin\theta d\varphi_1,\,
\sin\theta\cos\varphi_1 d\varphi_2,\,
\sin\theta\cos\varphi_1 \cos\varphi_2 d\varphi_3
)~,  \nonumber \\
\Omega^5{}_6&=&
-\sin\theta d\psi~,~~
\Omega^7{}_6=\cos\theta d\varphi_1~,~~
\Omega^8{}_6=\cos\theta \cos\varphi_1d\varphi_2~,~~
\Omega^8{}_7=-\sin\varphi_1 d\varphi_2~,
\nonumber\\
\Omega^9{}_6&=&\cos\theta \cos\varphi_1 \cos\varphi_2 d\varphi_3~,~~
\Omega^9{}_7=-\sin\varphi_1\cos\varphi_2 d\varphi_3~,~~
\Omega^9{}_8=
-\sin\varphi_2d\varphi_3~. \nonumber 
\end{eqnarray}

\section{Fermionic fluctuations of Wilson loop}

A supersymmetrized Wilson loop, which is proposed in \cite{DGO}, is
given by
\begin{eqnarray}
W = \frac{1}{N}{\rm Tr} P \left[
\e^{\int\bar{\zeta}(s) Q ds}\e^{\int 
(i A_{\mu}\dot{x}^{\mu} + \phi_i\dot{y}^i)ds } \e^{-\int\bar{\zeta}(s)Q
ds}
\right]\,.\nonumber 
\end{eqnarray}
Here the loop includes a superpartner of $(x^{\mu}(s),y^i(s))$, 
which couples to the fermion $\Psi$\,. 
The supersymmetry transformation in $\mathcal{N}$=4 SYM is given by 
\begin{eqnarray}
[Q,A_M]=\frac{i}{2}\Gamma_{M}\Psi\,, \qquad 
\{Q,\Psi\} = -\frac{1}{4}\Gamma_{MN}F^{MN}\,, \nonumber 
\end{eqnarray}
and the following relation may also be included:  
\[
 [Q,\dot{x}_M] = \frac{i}{4}\Gamma_M\dot{\zeta}\,.
\]

A small deformation for the fermionic variables may be considered. 
By setting that $\zeta=\bar{\zeta}=0$ on $C_0$\,, the only
contribution is evaluated as 
\begin{eqnarray}
\delta W|_{\rm fermion} = \frac{1}{N}\int\!ds{\rm Tr} P \left[
\delta\bar{\zeta}(s)R\Gamma_{\theta}\left(
1-i\Gamma_{\theta 6}
\right)
\Psi 
\e^{\int 
(i A_{\mu}\dot{x}^{\mu} + \phi_i\dot{y}^i)ds } 
\right]\,. \nonumber 
\end{eqnarray} 
Here note that the matrix defined as
\[
 \mathcal{P} \equiv \frac{1}{2}\left(1-i\Gamma_{\theta 6}\right) 
\]
is a projection operator. The original fermionic variable $\Psi$ has 16
components but it projects out half of it. As a result, the physical eight
components of the fermionic variables remain. 

Thus the operator insertion for the fermionic fluctuations are described
by the eight fermionic variables. We have discussed the circular case so
far, but the argument for the straight line is the same.  It is also the
same even for the BMN case, where the eight fermions $i\Gamma_1 h_+\Psi$
($h_+=\frac{1}{2}(1+i\Gamma_{12})$) are inserted.

\end{document}